\documentclass[twocolumn,
 reprint,
 amsmath,amssymb,
 aps,
]{revtex4-2}

\usepackage{graphicx}
\usepackage{dcolumn}
\usepackage{bm}
\usepackage[driverfallback=dvipdfmx,colorlinks,linkcolor=blue,citecolor=blue,urlcolor=blue]{hyperref}

\begin{document}

\title{6 GHz hyperfast rotation of an optically levitated nanoparticle in vacuum}

\author{Yuanbin Jin$^{1,2}$, Jiangwei Yan$^{1,2}$, Shah Jee Rahman$^{1,2}$, Jie Li$^{3}$, Xudong Yu$^{1,2,\dag}$, Jing Zhang$^{1,2,\ddag}$}
 \affiliation{$^1$The State Key Laboratory of Quantum Optics and Quantum Optics Devices, Institute of Opto-Electronics, Shanxi University, Taiyuan 030006, China \\
 $^2$Collaborative Innovation Center of Extreme Optics, Shanxi University, Taiyuan 030006, China \\
 $^3$Zhejiang Province Key Laboratory of Quantum Technology and Device, Department of Physics and State Key Laboratory of Modern Optical Instrumentation, Zhejiang University, Hangzhou 310027, China}

\begin{abstract}
We report an experimental observation of a record-breaking ultra-high rotation frequency about 6 GHz in an optically levitated nanoparticle system. We optically trap a nanoparticle in the gravity direction with a high numerical aperture (NA) objective lens, which shows significant advantages in compensating the influences of the scattering force and the photophoretic force on the trap, especially at intermediate pressure (about 100 Pa). This allows us to trap a nanoparticle from atmospheric to low pressure ($10^{-3}$ Pa) without using feedback cooling. We measure a highest rotation frequency about 4.3 GHz of the trapped nanoparticle without feedback cooling and a 6 GHz rotation with feedback cooling, which is the fastest mechanical rotation ever reported to date. Our work provides useful guides for efficiently observing hyperfast rotation in the optical levitation system, and may find various applications such as in ultrasensitive torque detection, probing vacuum friction, and testing unconventional decoherence theories.
\end{abstract}

\maketitle

\section{Introduction}

In recent years, levitated nanoparticles in vacuum have attracted considerable interests and become an important platform for ultrasensitive force detection~\cite{Geraci,Geraci2,Lukas16}, the study of macroscopic quantum phenomena~\cite{OMrmp,Arndt14,Markus20,millen20,klaus20}, and nonequilibrium thermodynamics~\cite{nonequil,nonequil1,nonequil2,millen1}, among many others. Over the past decade, significant progress has been made in the experimental realization of cooling the motion of trapped nanoparticles ~\cite{Raizen,Lukas12,Markus13,Arndt13,Barker15,Reimann,Markus19,sun,Lukas20} and the motional quantum ground state has been achieved~\cite{Markus20}. Such a system has also been employed for the fundamental test of unconventional decoherence theories at the macro scale~\cite{Bassi,Oriol,Jie,Barker16,Vinante,Du,Barker20}. In Refs.~\cite{Oriol,Jie,Barker16,Vinante,Du,Barker20} the relevant degree of freedom of motion is the center-of-mass (CoM) motion. Other degrees of freedom of motion of the levitated nanoparticle, such as the torsional vibration~\cite{Li16}, the precession motion~\cite{Rashid}, and rotation~\cite{Arita,Kuhn,Moore,Lukas18,Li18,Li20}, provide also rich physics to explore. Recent theoretical work~\cite{Hornberger,Bassi2} show that the rotational degree of freedom may offer considerable advantages in testing the continuous-spontaneous-localization collapse theory. Furthermore, hyperfast rotation~\cite{Lukas18,Li18,Li20} has many important applications, such as in testing material properties in extreme conditions~\cite{Schuck} and detecting the quantum form of rotational friction~\cite{Pendry}. Recently, a hyperfast rotation of frequency 5.2 GHz (700 MHz) of a trapped nanodumbbell (nanosphere) has been reported~\cite{Li20}. The rotation of a nanodumbbell is much faster than that of a nanosphere in the same size because it receives a much larger optical torque under the same trap and air pressure.

Stable optical levitation at low and high vacuum can be achieved without feedback cooling of the micro and nano-particle's motion. However, feedback cooling of the CoM motion is typically required to prevent particle loss from the trap at intermediate pressure (around 100 Pa), where photophoretic forces, sphere de-gassing, and other sources of noise not present in high vacuum may play significant roles~\cite{Monteiro}. In previous works, the hyperfast rotation~\cite{Lukas18,Li18,Li20} is observed for a horizontally propagating trapping beam with using feedback cooling to low pressure. A vertically-propagating trapping beam with a low NA is adopted to levitate the micrometer-sized spheres~\cite{Moore1}, in which feedback cooling is also used to reach low pressure and MHz rotation is obtained~\cite{Moore}. In this work, we show that, by adopting a vertical-up layout of the trapping beam and using a high NA objective lens to focus this laser, we can stably trap a nanoparticle from an atmospheric pressure to high vacuum ($10^{-3}$ Pa) without using feedback cooling. Therefore, our work could enable feedback-free optical trapping over almost full ranges of vacuum pressures. Consequently, we measure a fastest 4.3 GHz rotation of the trapped nanoparticle without feedback cooling. Nevertheless, for the fast rotation at high vacuum, the kinetic energy transfer between the rotation motion and the CoM motion makes the trap unstable ~\cite{Lukas18}, and hence the nanoparticle is easily lost from the trap. We thus apply the feedback cooling to the CoM motion, which improves the stability of the trap and makes it possible to reach even higher
vacuum, and consequently, we measure a highest rotation frequency about 6 GHz at $8\times10^{-3}$ Pa. This is, to our knowledge, the highest rotation frequency ever reported for a mechanical object.

\section{Trap nanoparticle from atmospheric to low pressure without using feedback cooling.}

\begin{figure}
\includegraphics[width=3in]{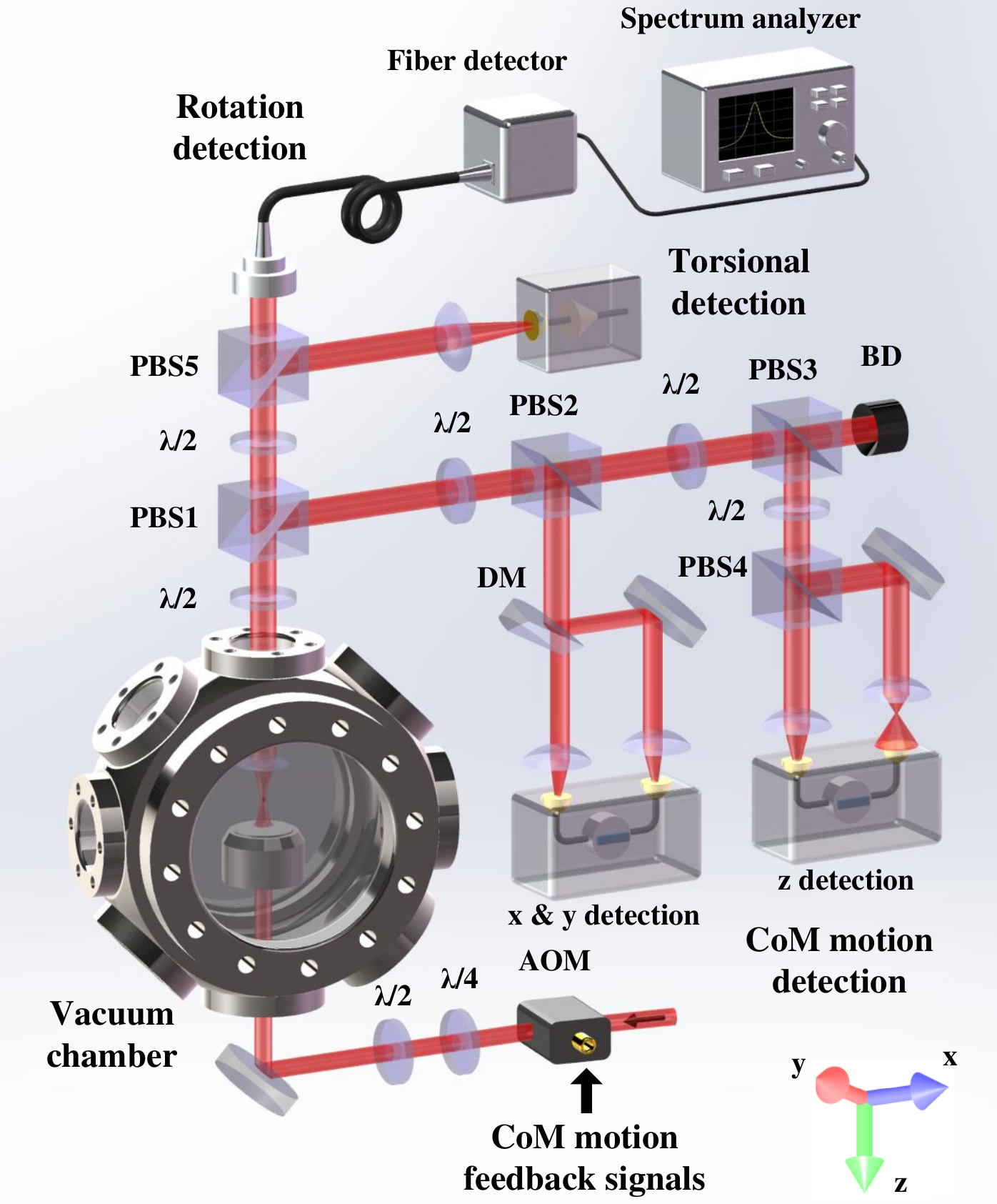}
\caption{(Color online). Schematic diagram of the experimental setup, which includes five parts: vacuum system, rotation detection, torsional vibration detection, CoM motion detection and feedback system. AOM: acousto-optic modulator; $\lambda/4$: quarter-wave plate; $\lambda/2$: half-wave plate; PBS(1-5): polarized beam splitter; DM: D-shape mirror; BD: beam dump.}
\label{fig1}
\end{figure}

The experimental setup is depicted in Fig.~\ref{fig1}. We optically trap an amorphous silica nanoparticle in vacuum using a TEM00 Gaussian mode 1064 nm laser in gravity direction. The laser first passes through an acousto-optic modulator (AOM) for shifting the frequency and controlling the power. The frequency-shifted laser is then coupled into a single-mode polarization maintaining fibre, of which the output beam passes successively through a
quarter and a half-wave plate. The vertically propagating 1064 nm laser is strongly focused by a high NA objective lens (Nikon CF IC EPI Plan 100X, the NA is 0.95 and the working distance (WD) is 0.3 mm) in a vacuum chamber for trapping the particles. The polarization of the light can be adjusted precisely by the combination of the two wave plates. The power of the laser before entering the chamber is 300 mW, and the total transmission of the chamber window and the objective lens is about $52\%$, leading to an effective trapping power about 156 mW in our experiment. The diameter of the trapping laser is 3.2 mm before entering the objective lens and about 1.1 $\mu$m at the focus point. The intensity distribution in the $x-y$ plane ($z$ in the axial direction) at the focus region is slightly asymmetric because of the vector diffraction of the light. The trapping light after the focus point is collimated by another high NA lens (Thorlabs C330TMD-C, NA=0.68, WD=1.8 mm) and then divided into three parts by two polarized beam splitters (PBS's). One part is used to measure the nanoparticle rotation signal using a detector (New focus 1554-A) with a flat gain of about $10^3$ V/A in a broad band range of DC-12 GHz, and the power input into the detector is 1 mW. The second part is detected by another broad band detector (Hamamatsu C12702-03 with a gain of about $10^4$ V/A in a bandwidth of DC-100 MHz) to measure the torsional vibration signal. The third part is used to measure the CoM motion in three directions. The CoM motion of the nanoparticle in the $x$ and $y$ directions are detected by using the D-shape reflective mirrors, which split the laser beams into two equal parts in space. The two parts are focused respectively by two short focus lenses ($f=30$ mm) and detected by a pair of photodiodes in the current-subtraction detectors. In order to detect the motion in the $z$ direction, the beam is separated by a beam splitter into two parts with the imbalanced intensity (1:2). One part is completely detected by the photodiode, while the other part is partially detected, but they are balanced in a current-subtraction detector. The balanced detectors have a high common mode rejection ratio (CMRR) and the gain is $10^4 V/A$. Those motional signals are finally analyzed by the spectrum analyzers.

A small dielectric particle in a strongly focused light beam feels a three-dimensional gradient force. In this situation, two relevant effects must be considered. First, for a single trapping beam configuration, the axial trapping force is crucial because the axial gradient force is small compared to the radial direction. Besides, in the axial direction, the particle also feels a scattering force from the light, which tends to push the particle out of the trap. Consequently, the equilibrium position of the particle is moved away from the focus point along the propagating direction of the trapping light, which decreases the well depth in this direction. Hence, for a horizontal layout of the trapping beam, a high NA lens is usually used to focus the beam and provides a large axial gradient force. As a result, the power density near the focus point is relatively high in this case. Second, in high vacuum the thermal transfer between the particle and the background gases is restrained. Therefore, the particle is heated to a high and uniform internal temperature. In parallel, in low vacuum the particle has a low and also uniform internal temperature due to a quick heat exchange between the nanoparticle and the air molecules. However, there are internal temperature gradients induced by the trapping laser at intermediate pressure, leading to a non-uniform distribution of temperature on the nanoparticle surface. When air molecules hit the nanoparticle, those rebounding from the warmer side will have higher energy than those rebounding from the cooler side. This imparts a net force (i.e., the photophoretic force) on the particle, which is in the vertical-up direction in our system. This force can easily kick the particle out of the trap at intermediate pressure~\cite{Monteiro}. Therefore, the feedback cooling of the CoM motion is generally required to stabilize the trap for the horizontal layout of the trapping beam because of the remarkable photophoretic force induced by the relatively large power density and the finite heat dissipation at intermediate pressure. Although a vertical-up layout for the trapping beam is also adopted in Ref.~\cite{Moore,Monteiro,Moore1}, in their experiment, the NA of the focusing lens is small, and thus the power density near the focus point is relatively low. Consequently, the well is shallow and the corresponding gradient force is weak. Therefore, the feedback cooling is also required to make their trap stable at intermediate pressure. To restrain these detrimental effects, we implement a vertical-up layout for the trapping light and use a high NA objective lens to strongly focus this beam , which can effectively compensate the influences of the scattering and photophoretic forces using its own gravity of the particle and simultaneously provide a large well depth. As a result, we can stably trap a nanoparticle from an atmospheric pressure to high vacuum without using feedback cooling. This results in about $50\%$ success probability of trapping a nanoparticle below an intermediate pressure 100 Pa to lower pressures. Furthermore, we can monitor the intensity of the scattering light from the trapping laser by imaging the nanoparticle via a charge-coupled device (CCD) camera. By further selecting the nanoparticles at atmospheric pressure with an intermediate scattering intensity, we can increase the success probability to more than $90\%$ below an intermediate pressure. Those nanoparticles with much higher or lower intensity of the scattering light cannot reach high vacuum in our experiment.

\begin{figure}
\includegraphics[width=3in]{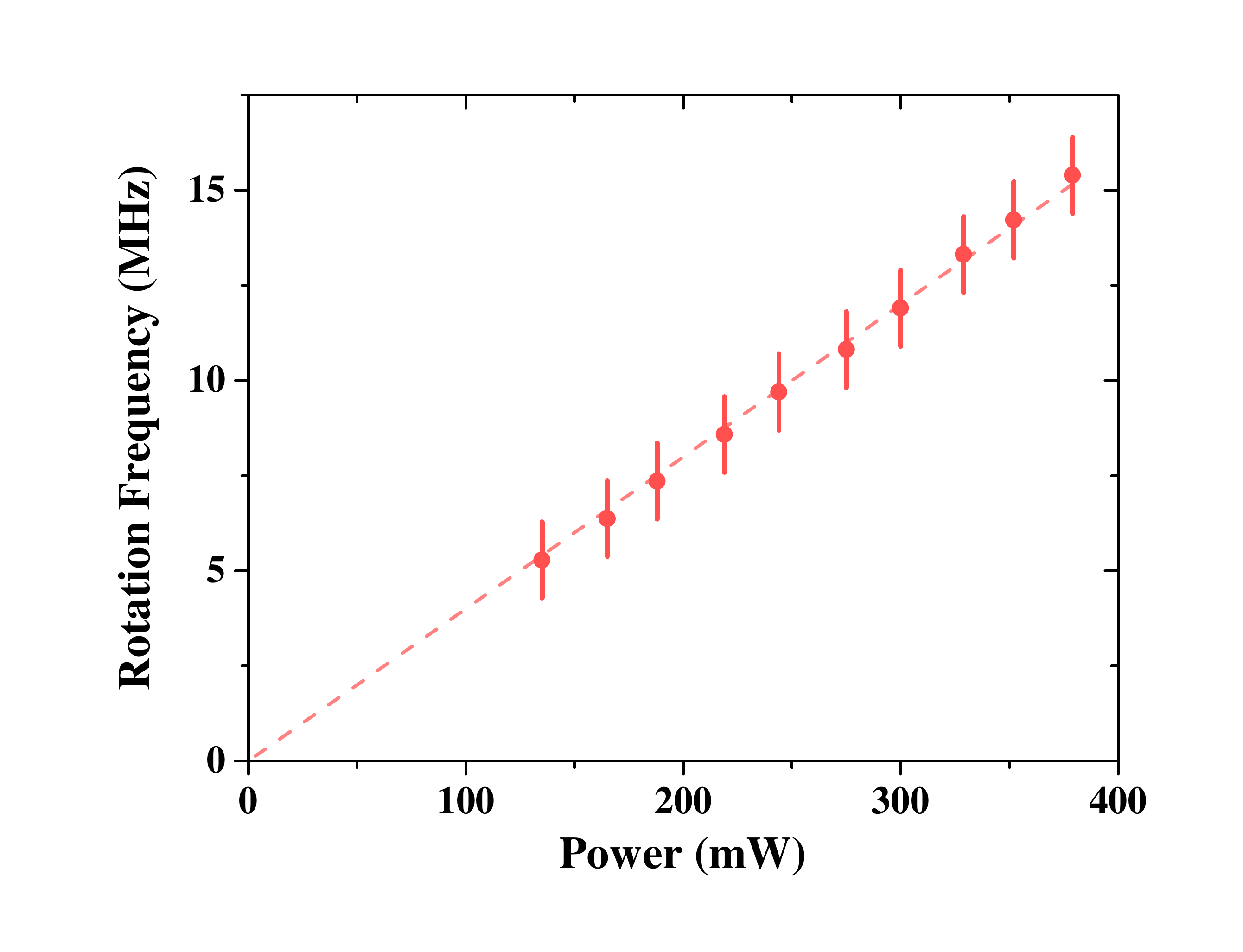}
\caption{(Color online). Measured rotation frequency versus the power of a near circularly polarized trapping laser at 1 Pa. The mass of the nanoparticle is about $7.2\times10^{-15}$ g. The dashed line is a linear fitting.}
\label{fig2}
\end{figure}

The trapped nanoparticle in the well can do translational motion, i.e., CoM motion, which generally has three eigen frequencies in three directions due to the vector diffraction of the light \cite{jin}. In our experiment, the eigen frequencies in $x$, $y$, $z$ directions for a linearly polarized trapping laser are about $f_{x}=200$ kHz, $f_{y}=240$ kHz, and $f_{z}=90$ kHz, respectively. Considering an ellipsoidal nanoparticle, the potential field aligns the longest axis of the particle along the linear polarization direction, and simultaneously leads to a torsion vibration~\cite{Li16,Li18,Li20}. By fitting the damping rates of the CoM's in the three directions (denoted as $\gamma_{x,y,z}$) and associating with the ratios of the damping rates at different pressures~\cite{Li18,Li20}, we can estimate the size and mass of the nanoparticle. The particles used in our experiment are not perfect spheres, and they are asymmetric. The measured average diameter is about $190$ nm, and the size difference of the three axes ($x$, $y$, $z$) is smaller than $60$ nm. For a linear polarization of the trapping beam (thus no rotation), we also test the stability of our system. We can stably trap the nanoparticles from air atmosphere to low pressure of $2\times10^{-3}$ Pa, which is ultimately limited by the air leakage of the vacuum chamber.

\begin{figure}
\includegraphics[width=3in]{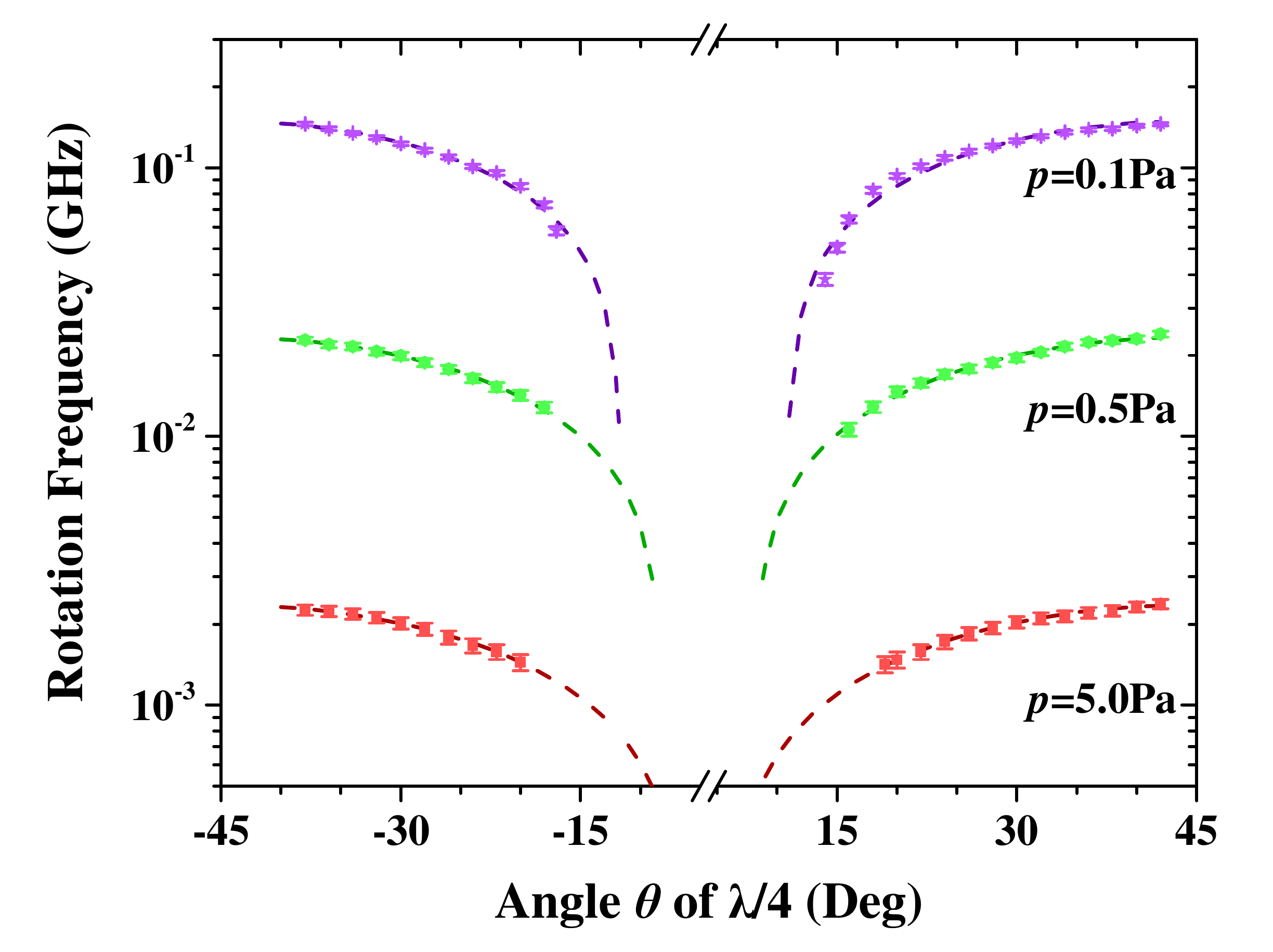}
\caption{(Color online). Measured rotation frequency versus the angle of the fast axis of the quarter-wave plate at different pressures. The red, green, and purple traces are measured at 5 Pa, 0.5 Pa and 0.1 Pa, respectively. The dash lines are the corresponding theoretical fittings using ${f_r} = {\mathop{\rm Re}\nolimits} [a\sqrt {{{\left( {1 - \cos b } \right)}^2}{{\sin }^2}\left( {2\theta} \right) - {{\sin }^2}\left( b  \right){{\cos }^2}\left( {2\theta} \right)} ]$ ~\cite{Friese1998,Lukas18}, where $\theta$ is the angle of the quarter-wave plate before the objective lens, $a$ depends on the air pressure and $b$ is the phase shift induced by the medium. The mass of the nanoparticle is about $6.5\times10^{-15}$ g.}
\label{fig3}
\end{figure}

\section{Rotation without using feedback cooling}

The angular momentum of the trapping light can be transferred to the nanoparticle due to the absorption, birefringence, and asymmetric shape of the particle \cite{Lukas18}. The transferred angular momentum provides a torque, which drives the particle to rotate. For an amorphous perfect nanosphere, the driving torque is only determined by the absorption of the doping impurity and the medium itself. For an imperfect nanosphere, the asymmetric shape can also induce a driving torque. We denote the total driving torque the particle receives as $M_{o}$. Meanwhile, the interaction with the gas molecules in the vacuum chamber damps the rotation of the particle, which causes a drag torque $M_{d}$. Under the driving and drag torques, the rotational motion equation of the particle is \cite{Lukas18}:
\begin{eqnarray}
2{\pi}I\frac{df_{r}}{dt}=M_{o}+M_{d},
\end{eqnarray}
where $I$ is the moment of inertia of the nanoparticle. The drag torque $M_{d}$ is proportional to the frequency of the rotation under a certain air pressure, $M_{d}=-2{\pi}If_{r}\gamma_{d}$~\cite{Frem}, where $\gamma_{d}=pR^2/({\eta}mv)$ is the damping rate of the rotation motion (which corresponds to the linewidth of the rotation signal, and $m$ is the mass of the nanoparticle), with $v$ the mean molecular velocity, and $\eta$ the accommodation factor accounting for the efficiency of the angular momentum transferred to the particle via collisions with gas molecules. According to this equation, in the beginning as the rotation gets faster under the driving torque, the drag torque increases accordingly. Eventually, the rotation speed increases to a certain point and remains constant at a certain air pressure as a result of the balance between the driving torque and the drag torque. The rotation frequency in the steady state can be solved, which is $f_{r}=\frac{1}{2{\pi}\gamma_{d}}\frac{M_{o}}{I}$. In order to measure the rotation frequency, the light after trapping the nanoparticle is split by a PBS and detected by a fast detector. Intuitively, a nanoparticle acts as a half-wave plate, and the rotated nanoparticle is like a polarization modulator. One period (2$\pi$) of the rotation of the nanoparticle will generate a 4$\pi$ modulation in the polarization of the trapping light. Thus, a frequency shift arises for the photons after interacting with the particle and the shift amount is $2f_{r}$, with $f_{r}$ the rotation frequency of the nanoparticle. Consequently, we obtain the $2f_{r}$ signal in the spectrum analyzer.

\begin{figure}
\includegraphics[width=3in]{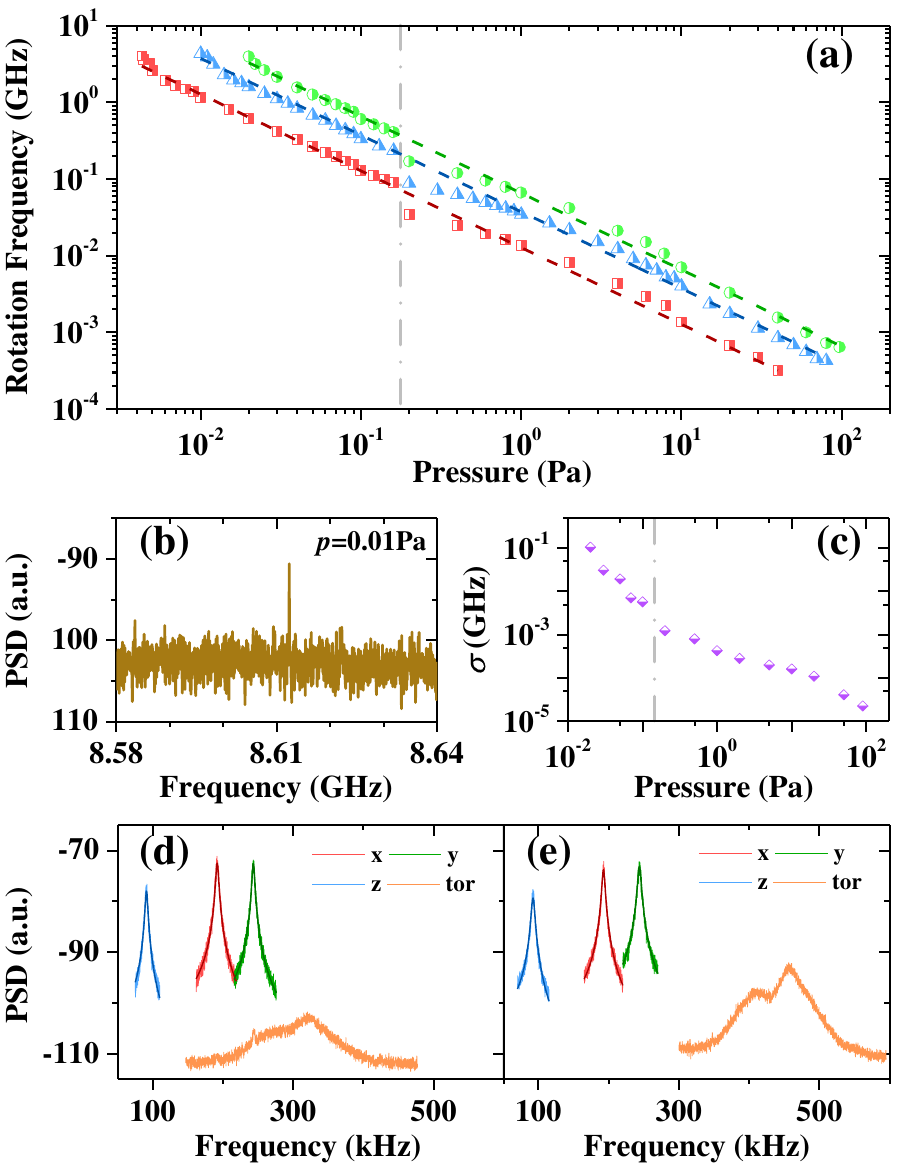}
\caption{(Color online). The experimental results without feedback cooling. (a) Measured rotation frequency of three trapped nanoparticles (green, blue, and red traces) versus air pressure without feedback cooling. The masses of the three nanoparticles are about $5.8\times10^{-15}$ g, $1.1\times10^{-14}$ g, and $7.2\times10^{-15}$ g, respectively. The dashed color lines are the corresponding theoretical fittings according to the inverse ratio of the rotation frequency and the pressure. (b) Power spectral density of a rotation signal of 8.6 GHz at 0.01 Pa. (c) Standard deviation of the rotation frequency versus air pressure measured for one of the nanoparticles. At each pressure, we perform 120 measurements to obtain the standard deviation. In (a)-(c), the measurements are performed using a near circularly polarized trapping laser with a fixed laser power 300mW. The gray dashed lines in (a) and (c) are the boundaries of the pressure measurements using the two gauges. (d) and (e) The CoM motion and torsional vibration signals of two nanoparticles (corresponding to red and green traces in (a)) at 500 Pa for a linearly polarized trapping laser (the polarization direction is along x axis). The damping rates of the CoM motions in three directions are different, which indicates that the nanoparticles are not perfect spheres~\cite{Li18,Li20}. In (d), $\gamma_{x}=2.69$ kHz, $\gamma_{y}=2.78$ kHz,  $\gamma_{z}=2.83$ kHz, thus the ratios are $\gamma_{y}/\gamma_{x}=1.03$ and $\gamma_{z}/\gamma_{x}=1.05$. In (e), $\gamma_{x}=2.34$ kHz, $\gamma_{y}=2.80$ kHz,  $\gamma_{z}=2.67$ kHz, thus the ratios are $\gamma_{y}/\gamma_{x}=1.20$ and $\gamma_{z}/\gamma_{x}=1.14$.}
\label{fig4}
\end{figure}

Considering the circularly polarized trapping laser, the total driving torque is proportional to the light intensity: $M_o \,{\propto} \, I_e$
($I_{e}$ is the intensity of the trapping light at the equilibrium point of the particle). Hence, the rotation frequency shows a linear dependence upon the trapping laser power. This is clearly seen in Fig. 2 for a near circularly polarized trapping beam. Moreover, the rotation direction can be altered by changing the chirality of the light. For the elliptical polarization, the light can be decomposed into a circular and a linear polarization component. The birefringence and asymmetric shape of the particle align the particle along the linear polarization and result in the torsional vibration, while the circular polarization component drives the particle to rotate \cite{shao}. Therefore, the weights of these two components determine the motion of the particle: If the effect of the circular polarization component is stronger than that of the linear polarization, the particle starts to rotate; if the opposite, the rotation would not occur and the torsional vibration can be observed. Here, the ellipticity of polarization is controlled by adjusting the angle of the fast axis of the quarter-wave plate. Rotation motion disappears in a certain angle range, as shown in Fig.~\ref{fig3}. As we change the chirality of the polarization, the rotation direction of the nanoparticle is changed.

In order to observe the rotation of the nanoparticle, we first trap the nanoparticle below an intermediate pressure 100 Pa to lower pressures with success probability more than $90\%$, and we then can observe the rotation in high vacuum with probability about $90\%$.  In Fig.~\ref{fig4} (a), we measure the rotation frequency of three trapped nanoparticle versus the air pressure for a fixed laser power 300 mW without feedback cooling. We use two vacuum gauges, a resistance gauge with measurement range from $5\times10^{-2}$ Pa to $10^5$ Pa, and a hot cathode ionization gauge with measurement range from $10^{-7}$ Pa to 0.2 Pa. This results in a slight mismatch between the two traces measured by the two vacuum gauges for the same nanoparticle at pressure around 0.2 Pa. We observe a beat signal of about 8.6 GHz, corresponding to a rotation frequency about 4.3 GHz, at 0.01 Pa, as shown in Fig.~\ref{fig4} (b).

The rotation and the CoM motion can influence each other, which leads to the coupling of the two motions. On the one hand, a large amplitude of the CoM motion causes a large change of the laser strength that the nanoparticle feels, and consequently induces a large fluctuation of the rotation frequency (the rotation frequency depends on the strength of the trapping laser). On the other hand, when $f_{r} \gg f_{CoM}$ ($f_{CoM}=f_{x,y,z}$) and the fluctuation of the rotation frequency $\sigma\geq f_{CoM}$, the energy of rotation can also be transferred to the CoM motion. Thus, the fluctuations of CoM and rotation motions are correlated. We can use an active feedback cooling of the CoM motion to minimize the torque fluctuation of the trapping laser for the nanoparticle, and reduce the standard deviation of the rotation frequency. In Fig.~\ref{fig4} (c), we show the fluctuation of the rotation frequency for one of the nanoparticles. The frequency uncertainty becomes larger as the pressure reduces.

\section{Rotation with feedback cooling}

\begin{figure}
\includegraphics[width=3in]{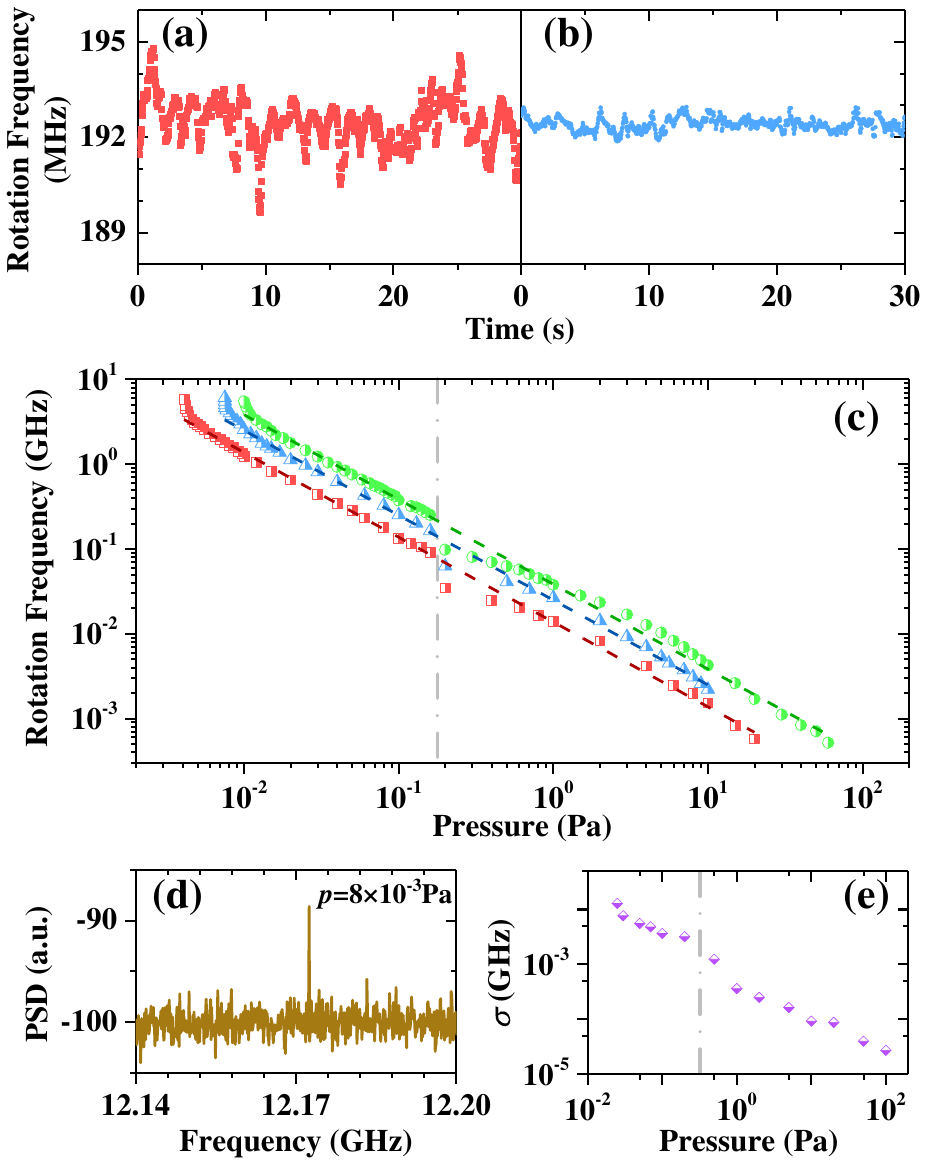}
\caption{(Color online). The fluctuation of the rotation frequency (a) without feedback cooling and (b) with feedback cooling at 0.16 Pa. We sample the rotation frequency 30 times per second. (c) Measured rotation frequency of three trapped nanoparticles (green, blue, and red traces) versus air pressure with feedback cooling. The masses of the three nanoparticles are about $1.2\times10^{-14}$ g, $7.4\times10^{-15}$ g, $7.6\times10^{-15}$ g, respectively. The dashed color lines are the corresponding fittings using the inverse ratio of the rotation frequency and the pressure. (d) Power spectral density of a rotation signal of 12.17 GHz at $8\times10^{-3}$ Pa. (e) Standard deviation of the rotation frequency versus air pressure measured for one of the nanoparticles. The gray dashed lines in (c) and (e) are the boundaries of the pressure measurements using the two gauges. The trapping beam is a near circularly polarized laser in these measurements.}
\label{fig5}
\end{figure}

In order to reduce those deleterious factors, we further implement feedback controls to cool the CoM motion of the nanoparticle in three directions (see Fig.~\ref{fig1}). The displacement signals in three directions are sent into broad-bandwidth lock-in amplifiers (Zurich Instruments HF2LI 50 MHz) for generating the corresponding double-frequency signals, which are then input into a function generator of AOM for modulating the power of the trapping laser and cooling the CoM motions \cite{Ulbricht17}. This parametric feedback cooling results in significantly improved stability of the trap. Figure~\ref{fig5}(a) and (b) illustrate the fluctuation of the rotation frequency before and after the feedback cooling at 0.16 Pa, respectively. Figure~\ref{fig5}(c) shows the rotation frequencies versus the air pressure with feedback cooling for three different nanoparticles. The highest rotation frequency observed is about 6 GHz at $8\times10^{-3}$ Pa and the corresponding beat signal is 12.17 GHz, as shown in Fig.~\ref{fig5} (d). At the rotation frequency higher than 4 GHz, the linear dependence of $\gamma_{d}$ on the pressure is no longer valid, which results in steeper slopes of the traces. In this region, the raising of the rotation frequency is very fast and the nanoparticle is easily lost as we further reduce the pressure. Therefore, we must carefully control the evacuating speed of the vacuum pump to obtain the highest rotation frequency. Fig.~\ref{fig5} (e) shows the fluctuation of the rotation frequency measured for one of the nanoparticles with feedback cooling. The fluctuation of the rotation frequency is significantly reduced by the feedback cooling, comparing with Fig.~\ref{fig4} (c). The different motion states for the respective nanoparticles in Fig.~\ref{fig4} and Fig.~\ref{fig5} are caused by their different properties, which originate from the polydispersity.

\section{Discussion}

In this paper, we have adopted a vertical-up layout of the trapping light and used a high NA objective lens to strongly focus this beam in an optical levitation system, which allows us to trap a nanoparticle from an atmospheric pressure to high vacuum without using feedback cooling. Consequently, we measure a maximum rotation frequency 4.3 GHz without feedback cooling. Apparently, without using feedback controls, the experiment will be more compact and cost-saving. Trapped nanoparticles in high vacuum promise many important studies, such as Refs.~\cite{Geraci,Geraci2,Lukas16,OMrmp,Arndt14,Markus20,millen20,klaus20,nonequil,nonequil1,nonequil2,millen1}. Therefore, our work (trapping without feedback controls) will find important applications in the above studies. By further including feedback cooling, we have measured a record high rotation frequency about 6 GHz. In our experiment, the rotation is hyperfast and close to the regime where the internal forces generated were strong enough to break up the material. Our work thus provides an important platform for studying vacuum friction and the material properties under extreme conditions. The system can also be used for ultrasensitive torque detection~\cite{Li20} and micron-scale pressure gauges~\cite{CPB}. Furthermore, our work sheds light on the test of the continuous-spontaneous-localization collapse theory by using the rotational degrees of freedom~\cite{Hornberger,Bassi2}.

\begin{acknowledgments}
Ministry of Science and Technology, China (Grant No. 2016YFA0301602), National Research Foundation of China (Grant No. 11234008, 61975101, 11474188, 11704234) and the support from the XPLORER PRIZE.
\end{acknowledgments}

\appendix

\section{Loading Process}

To load the nanoparticle, we tried amorphous silica nanoparticles produced by different manufacturers, and finally selected the non-functionalized silica nanoparticle (Bangs Laboratories, Inc.), which gave us the best result. Its nominal diameter is about 170 nm with the range of 20$\%$. The hydro-soluble silica nanoparticles are first diluted in the high-purity ethanol with concentration of about $1.5\times 10^7/ml$ and are then sonicated for 30 minutes. The dilution solution is poured into an ultrasonic nebulizer (OMRON NE-U22). The droplets containing the nanoparticles are dispersed by the ultrasonic nebulizer and guided through a thin tube near the focus of the objective lens in the vacuum chamber. Once a particle is trapped in the focused beam, the vacuum pump then starts to evacuate the chamber.

\nocite{*}

\bibliography{references}

\end{document}